\documentclass[pra,aps,twocolumn,showpacs,superscriptaddress]{revtex4-1}
\pdfoutput=1
\usepackage{amsmath,amssymb,bm,amsthm,bbold}
\usepackage[final]{graphicx}
\usepackage[colorlinks=true,citecolor=blue,linkcolor=blue,urlcolor=blue]{hyperref}
\usepackage[T1]{fontenc}
\usepackage{mathptmx}

\usepackage{mathrsfs}

\usepackage{color}
\definecolor{ForestGreen}{rgb}{0.3,0.6,0.3}

\usepackage[LowerCaseTrace,SquareTraceBrackets]{quantum}

\newcommand{\ud}{\mathrm{d}}
\newcommand{\la}{\left\langle}
\newcommand{\ra}{\right\rangle}

\renewcommand{\one}{\mathnormal{1}}

\newcommand{\modsq}[1]{\Abs{#1}^2}

\begin{document}

\title{Parameter estimation using NOON states over a relativistic quantum channel}

\author{Dominic Hosler}\email{dominichosler@physics.org}
\affiliation{Department of Physics and Astronomy, The University of Sheffield, S3 7RH, United Kingdom}

\author{Pieter Kok}\email{p.kok@sheffield.ac.uk}
\affiliation{Department of Physics and Astronomy, The University of Sheffield, S3 7RH, United Kingdom}

\begin{abstract}
\noindent
We study the effect of the acceleration of the observer on a parameter estimation protocol using NOON states. An inertial observer, Alice,  prepares a NOON state in Unruh modes of the quantum field, and sends it to an accelerated observer, Rob. We calculate the quantum Fisher information of the state received by Rob. We find the counterintuitive result that the single-rail encoding outperforms the dual rail. The NOON states have an optimal $N$ for the maximum information extractable by Rob, given his acceleration. This optimal $N$ decreases with increasing acceleration.
\end{abstract}


\date{\today}
\maketitle

\section{Introduction}
Metrology is the study of precision measurements and exactly how precise the measurement of a physical quantity can be.
Any physical measurement of a continuous parameter can be thought of in three stages: First, we prepare an initial quantum state; second, the quantum state evolves to a new state that depends on the parameter of interest; and third, a suitable measurement of the system will reveal information about the parameter. Repeating the three steps of the measurement process $N$ times independently, reduces the uncertainty in the parameter $\theta$ by a factor $\delta \theta = 1/\sqrt{N}$. This is the shot-noise limit, or the standard quantum limit (SQL). It can be surpassed by exploiting quantum effects \cite{Caves1981,Yurke1986,Holland1993,Bollinger1996}, leading to an ultimate quantum scaling of $\delta \theta = 1/N$. This is the optimal precision for parameter estimation in quantum mechanics, and it is known as the Heisenberg limit \cite{Holland1993,Giovannetti2006,Zwierz2010,Giovannetti2011}.  Suggestions have been made to use these quantum improvements for a superior signal-to-noise ratio when reading from classical optical storage \cite{DallArno2013}, and for the measurement of position and orientation via interferometers and interferometric gyroscopes \cite{Dowling1998,Barnett2003}, among other things.

We study the situation where an inertial observer, Alice, prepares optical systems in highly nonclassical states and subjects them to a dynamical evolution that imparts a relative phase $\theta$. If the optical system consists of two modes (i.e., dual-rail encoding), the NOON state is chosen:
\begin{align}\label{eq:dualrailstate}
 \ket{\Psi}=\frac{\ket{N,\varnothing} + e^{i N\theta}\ket{\varnothing,N}}{\sqrt{2}}\, , 
\end{align}
where $\ket{\varnothing}$ is the vacuum state and $\ket{N}$ is the Fock state of $N$ photons in a particular mode. Alternatively, Alice may prepare a single-mode optical system in a state similar to the NOON state:
\begin{equation}
  \label{eq:singlerailstate}
  \ket{\psi}= \frac{\ket{\varnothing} + e^{i N \theta}\ket{N}}{\sqrt{2}}\, .
\end{equation}
We refer to this state as the single-rail NOON state. In an ideal situation these NOON states are optimal for measuring the phase $\theta$ in an interferometer \cite{Kapale2005}. The minimum uncertainty achievable by a suitable quantum measurement on these NOON states is proportional to the Heisenberg limit $1/N$. Alice sends the optical system to a uniformly accelerating observer, Rob, who measures the phase $\theta$. He will experience Unruh radiation as a consequence of his acceleration  \cite{Hawking1975,Unruh1976,Takagi1986}, and this will deteriorate the quantum states he receives from Alice. We calculate the quantum Fisher information about $\theta$ as seen by Rob, which (via the Cram\'er-Rao bound \cite{Kok2010}) is an indication of the smallest phase change $\delta\theta$ that he can detect.

Noise and errors are an inescapable part of any experiment, and quantum metrology must take into account various different types of noise and compensate if possible. In an interferometer, losses and imperfect measurements can reduce the precision of measurements made. In many cases the losses mean that quantum effects can result only in a constant factor improvement over the classical bounds \cite{Kolodynski2010,Ono2010}. There are modifications that can be made to the input state to minimise the damage from these losses \cite{Dobrzanski2009}. Other types of noise can sometimes still allow an asymptotic improvement over the classical bounds \cite{Chaves2012}. Decoherence noise becomes a problem when measuring phases using an interferometer, but there are methods to overcome this \cite{Matsuzaki2011,Genoni2011}. While in the presence of decoherence the Heisenberg limit may not be achievable, improvement over the shot noise limit is still possible \cite{Dobrzanski2012}.  We show that in the case of communicating NOON states with an accelerated observer, there exist an optimal $N$ for detecting phase changes, beyond which the precision deteriorates.

\section{Quantum Metrology}
Every quantum measurement of a continuous parameter $\theta$ that is not itself associated with a Hermitian operator must take place via the measurement of some observable $A$. The relationship between $A$ and $\theta$ must be known, in order to extract the value of $\theta$ from the experiment. If translations in $\theta$ are generated by a Hermitian operator $K$, the Robertson relation for the variances in $A$ and $K$ gives us
\begin{align}
 \Delta A \Delta K \geq \frac12 \left| \braket{[A,H]} \right|\, .
\end{align}
Using the generalised Heisenberg equation 
\begin{align}
 \frac{d {A}}{d\theta} = \frac{1}{i\hbar} {[A,K]} \, , 
\end{align}
we can write the uncertainty relation as $\delta\theta \Delta K \geq \frac12 \hbar$, with 
\begin{align}
 \delta\theta \equiv \Delta A\; \left| \frac{d\la A\ra}{d\theta} \right|^{-1}\, ,
\end{align}
where $\la A\ra$ is the expectation value of $A$. The uncertainty $\delta\theta$ is lower bounded by the Cram\'er-Rao bound \cite{Braunstein1994,Kok2010}
\begin{align}
  (\delta \theta)^2 \geq \frac{1}{N \mathscr{F}(\theta)},
\end{align} 
where $\mathscr{F}(\theta)$ is the Fisher information and $N$ is the number of independent measurements.
This bounds the achievable precision on $\theta$ for any physical measurement and can be saturated by a well-chosen measurement procedure.
The Fisher information can be thought of as the maximum amount of information we can learn about the parameter in any single measurement.
If $\mathscr{F}$ is constant, the Cram\'er-Rao bound reduces to the SQL.
On the other hand, if the resources of all $N$ measurements are put together in an entangled state that is subjected to a single measurement, the Fisher information can be as large as $\mathscr{F} = N^2$.
This is the case for NOON states. 

The Fisher information depends on the state of the system and the observable that is measured.
The Cram\'er-Rao bound gives the maximum precision in $\theta$ for a \emph{particular} measurement scheme.
On the other hand, we can choose various observables that may lead to different Fisher information.
So instead we ask what is the information that is intrinsic in the quantum state, without any reference to the actual measurement procedure.
This is called the \emph{quantum} Fisher information, and it is at least as large as the Fisher information for the optimal observable $A$.
The Cram\'er-Rao bound is also saturated for the quantum Fisher information \cite{Braunstein1994}.

The quantum Fisher information is calculated from the density matrix of the state,
\begin{equation}
  \label{eq:fischerinfo}
  \mathscr{F}(\theta)= \Tr{{\rho'} \mathscr{L}_\rho({\rho'})} \qquad\text{with}\qquad \rho' = \frac{\ud \rho}{\ud \theta},
\end{equation}
and $\mathscr{L}_{\rho}$ is the symmetric logarithmic derivative given by
\begin{equation}
  \label{eq:loweringoperator}
  \mathscr{L}_\rho(B)=\sum_{jk} \frac{2B_{jk}}{p_j+p_k} \Ketbra{j|k} \junktofixketbra,
\end{equation}
where the $p_j$ are the eigenvalues of $\rho$, $\{\ket{j}\}$ is the eigenbasis of $\rho$, and $B$ is an arbitrary self-adjoint operator in the same space as $\rho$.
Calculating the quantum Fisher information for states passing through a quantum channel will allow us to determine the fundamental limit to the precision in $\theta$, and it will tell us about the information preserving capabilities of the various channels.

\begin{figure}[tb]
  \begin{center}
      \includegraphics[width=0.8\columnwidth]{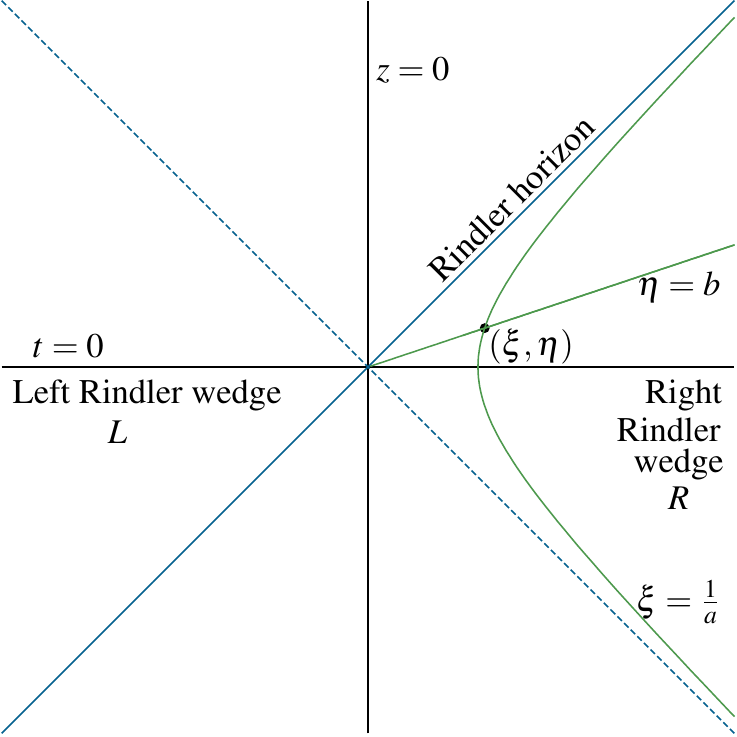}
  \caption[Minkowski diagram of Rindler coordinates]{A Minkowski space-time diagram separated into the left and right wedges and showing the Rindler coordinates. Rob travels along a trajectory with fixed spatial Rindler coordinate $\xi=\frac1a$. Alice has the freedom to create excitations in one or both of the wedges. In this paper, Alice only creates excitations in the right wedge, corresponding to the SWM.}
  \label{fig:rindler-coords}
  \end{center}
\end{figure}

\section{Relativistic noise} 
The observer Rob is accelerating with uniform proper acceleration $a$.
The task is for Rob to measure the field and obtain an estimate for the parameter $\theta$.
Rob's detector will be subject to the relativistic noise present in the form of the Unruh-Hawking effect \cite{Hawking1975,Unruh1976,Takagi1986}.
Rob cannot describe the entire field as seen by Alice because he is in the presence of a horizon.
We require an additional observer, anti-Rob, to live behind the horizon.
Rob and anti-Rob both describe their regions of the field by Rindler modes.
Rob's region is denoted with a subscript $R$, and anti-Rob's with a subscript $L$, corresponding to the right and left wedges of a Minkowski space-time diagram (see Fig.\nobreakspace \ref {fig:rindler-coords}).
Alice creates the single and dual-rail NOON states, in particular Unruh modes.
These modes are used because they have a one-to-one correspondence with Rindler modes, which in turn map to Rob's observed optical modes.

The Rindler modes are a family of complete sets of mode functions; each value for Rob's acceleration constructs a different set of Rindler modes.
This means we must choose the Unruh modes for Alice, depending on the value of Rob's acceleration.
The Unruh modes are not physically realisable due to highly oscillatory behaviour close to the horizon from an inertial perspective.
They also are fully delocalised, which makes them only partially measurable using local measurements.
We use the Unruh modes because they form a complete orthonormal set and they have been conjectured to provide the limiting case for informational quantities.

As an idealisation, Alice is assumed to have access to any part of Minkowski space, and she can therefore create any superposition of modes in any part of her space-time.
The Unruh modes map to single frequency Rindler modes with support in both Rob's and anti-Rob's regions of space-time.
This is described by the transformation of creation operators, $\hat{a}^\dagger_P = q_L \hat{A}^\dagger_L + q_R \hat{A}^\dagger_R$, where $q_L$ and $q_R$ are complex numbers such that $\modsq{q_L}+\modsq{q_R}=1$ \cite{Bruschi2010}.
In this paper we use the so-called single wedge mapping (SWM), in which $q_R=1$ and $q_L=0$
\footnote{The SWM has been called the Single Mode Approximation in the literature, which is has misleading connotations in quantum optics. The terminology SMW makes it clear that we are creating excitations in Rob's space-time wedge only.}.

The transformation between Unruh modes and Rindler modes is well known \cite{Takagi1986}.
It takes the form of two mode squeezing between the equivalent modes in Rob and anti-Rob's regions
\begin{equation}
 \label{eq:twomodesqueezing}
\widehat{U} = \exp\left[i r \hat{a}_R \hat{a}_L - i r \hat{a}_R^\dagger \hat{a}_L^\dagger \right].
\end{equation}
The squeezing parameter $r$ \footnote{The Bogoliubov transformation represented by \mbox{$\widehat{U}$} is a squeezing operation in the phase space of the modes \mbox{$\hat{a}_R$} and \mbox{$\hat{a}_L$}.} is related to the acceleration of Rob and the frequency of the Rindler mode by
\begin{equation}
  \label{eq:squeezingparameter}
  \tanh r = \exp \left(\frac{- \omega \pi}{a} \right),
\end{equation}
where $\omega$ is the frequency of the mode and $a$ is Rob's acceleration.
Once the modes in the region behind the horizon (corresponding to anti-Rob) are traced out, we find that Rob is left with a mixed state that includes thermal noise. We want to know the amount of information Rob can extract from the NOON states when this noise is present.

Using the dual- and single-rail NOON states given in Eqs.~(\ref {eq:dualrailstate}) and (\ref {eq:singlerailstate}), the parameter $\theta$ is encoded in the relative phase.
We calculate the quantum Fisher information of the NOON states in Rob's accelerated frame.
To this end, we use the transformation operator defined in Eq.\nobreakspace \textup {(\ref {eq:twomodesqueezing})}.
When applied to a single mode with $N$ excitations, the joint state held by Rob and anti-Rob is
\begin{align}
  \label{eq:transformedN}
  \ket[R\bar{R}]{\psi} = \sum_{p=0}^\infty \frac{(ir)^p}{|r|^p} \frac{\tanh^p |r|}{\cosh^{N+1}|r|} \begin{pmatrix}p+N \\ p\end{pmatrix}^\frac12 \nonumber \\
  \ket[R]{N+p} \otimes \ket[\bar{R}]{p}.
\end{align}
To transform the dual- and single-rail NOON states, we construct Alice's density matrix from the states with zero and $N$ excitations, then transform each term according to Eq.\nobreakspace \textup {(\ref {eq:transformedN})}.
Finally, we trace out anti-Rob leaving Rob with a mixed state represented by an infinite matrix in the number basis.

We have not found an analytic expression for the quantum Fisher information, and to calculate the quantum Fisher information numerically we need a finite matrix.
The coefficients of the matrix decay as the number of excitations increases, and we can therefore use a truncation of the matrix to get the result within a specified precision.
We numerically evaluated the matrices, for a size $k$, and calculated the quantum Fisher information.
To check that the truncation was sufficient, we calculated the quantum Fisher information for matrices of size $k+1$.
We continued to increase $k$ until the quantum Fisher information changed less than the prescribed absolute precision of $10^{-5}$.

Using this truncated matrix $\rho$, we calculated the quantum Fisher information using the algorithm as follows:
(i) find eigenvalues and eigenvectors of $\rho$;
(ii) differentiate $\rho$ with respect to $\theta$ to find $\rho'$;
(iii) use the eigenvectors of $\rho$ to transform both $\rho$ and $\rho'$ into the basis where $\rho$ is diagonal;
(iv) perform the lowering operator according to Eq.\nobreakspace \textup {(\ref {eq:loweringoperator})}; and
(v) calculate the quantum Fisher information using the trace of the matrix product of $\rho'$ and the lowered $\rho'$.
The quantum Fisher information includes $\theta$ in its calculation, however, the final result does not depend on $\theta$. The quantum Fisher information does depend on both the squeezing parameter $r$, and the number $N$ of excitations Alice created in the state.

\begin{figure}[tb]
  \begin{center}
      \includegraphics[width=\columnwidth]{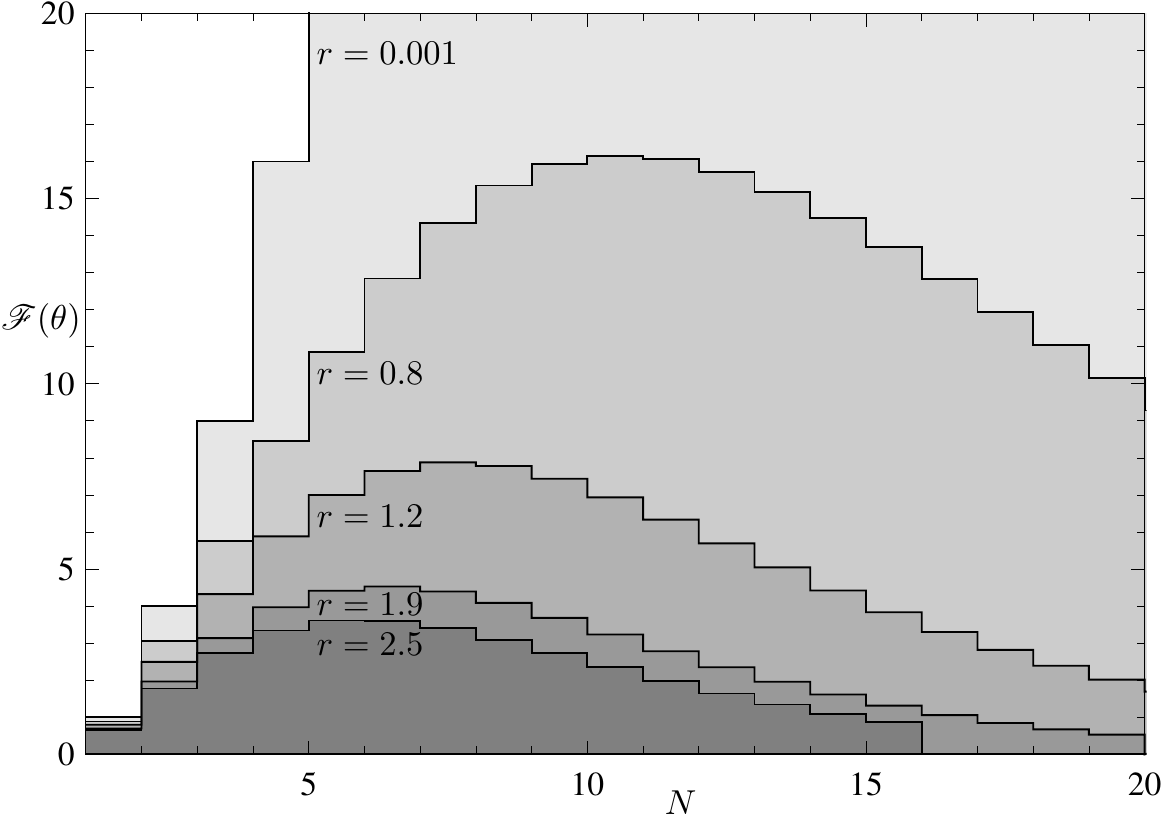}
  \caption[Single rail quantum Fisher information over $N$]{The quantum Fisher information $\mathscr{F}(\theta)$ over $N$ for Rob measuring the parameter $\theta$ when Alice has sent him the state using the single rail for various values of $r$.}
  \label{fig:single-noon-N}
  \end{center}
\end{figure}

\begin{figure}[tb]
  \begin{center}
      \includegraphics[width=\columnwidth]{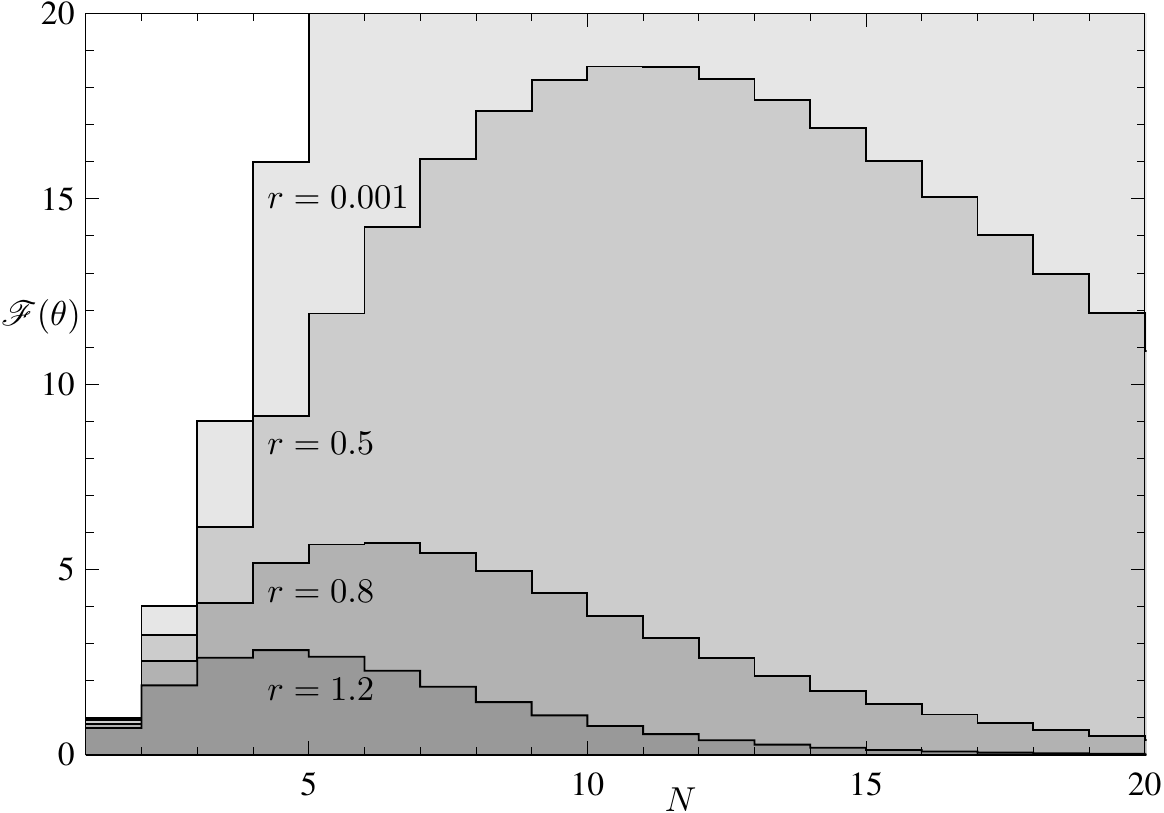}
  \caption[Dual rail quantum Fisher information over $N$]{The quantum Fisher information $\mathscr{F}(\theta)$ over $N$ for Rob measuring the parameter $\theta$ when Alice has sent him the state using the dual rail for various values of $r$.}
  \label{fig:dual-noon-N}
  \end{center}
\end{figure}

\section{Results and discussion}

\begin{figure}[t]
  \begin{center}
      \includegraphics[width=\columnwidth]{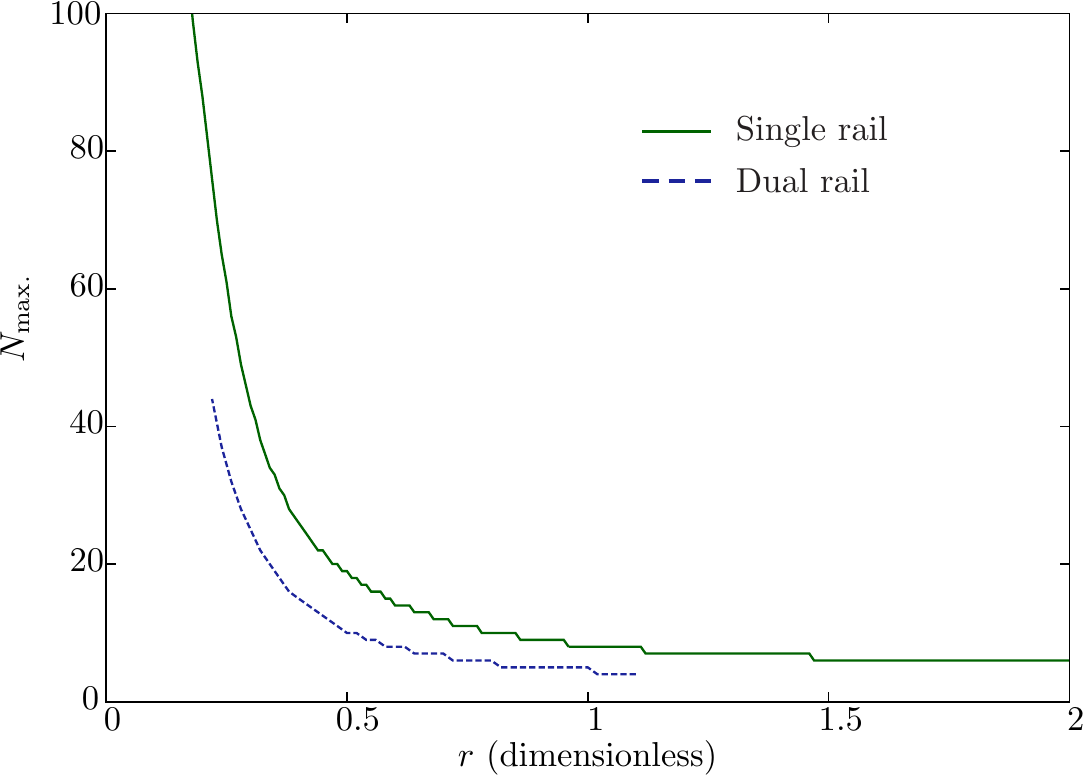}
  \caption[Value of $N$ for maximum quantum Fisher information over $r$]{The value of $N$ resulting in a maximum quantum Fisher information $\mathscr{F}(\theta)$ for Rob measuring the parameter $\theta$ when Alice has sent him the state using the single rail over $r$ for both the single rail and dual rail. The dual rail has a shorter line because the computational resources required prohibited further calculation in either direction.}
  \label{fig:single-noon-maxN}
  \end{center}
\end{figure}

The quantum Fisher information as a function of $N$ for the single-rail encoding is shown in Fig.\nobreakspace \ref {fig:single-noon-N}, and the dual-rail encoding is shown in Fig.\nobreakspace \ref {fig:dual-noon-N}. In both cases the channel is similar to an amplifying channel, in that the more excitations that are present initially, the more excitations are created by the channel. The excitations created by the channel are thermal, and more initial excitations contributes to a higher amount of noise. For a fixed acceleration we find an exponential decay in the quantum Fisher information with respect to $N$. This results in an optimal value of $N$ for each particular noise level, which is in turn determined by the acceleration $a$ and the Unruh frequency $\omega$.
Higher $N$ states sent via the dual rail are more susceptible to the noise than the single-rail equivalents, due to the fact that both the logical zero and logical one contain $N$ excitations, whereas in the single rail, the ``zero'' state contains no excitations. This results in a smaller expected number of excitations for the single rail, and hence less noise.

Previous research has shown that the dual-rail channel tends to perform better than the single-rail channel for quantum communication between inertial Alice and accelerated Rob \cite{Hosler2012,Martin-Martinez2012a}. Here we find the opposite effect, that the single-rail channel performs better. The reason for the difference is that these studies are measuring a different quantity. The information in the states studied previously was contained in the relative amplitude of the logical zero and one basis states. By contrast, in this study, the information about $\theta$ is contained entirely in the relative phase. The preference for single- or dual-rail encoding therefore depends critically on the communication task.

\begin{figure}[t]
  \begin{center}
      \includegraphics[width=\columnwidth]{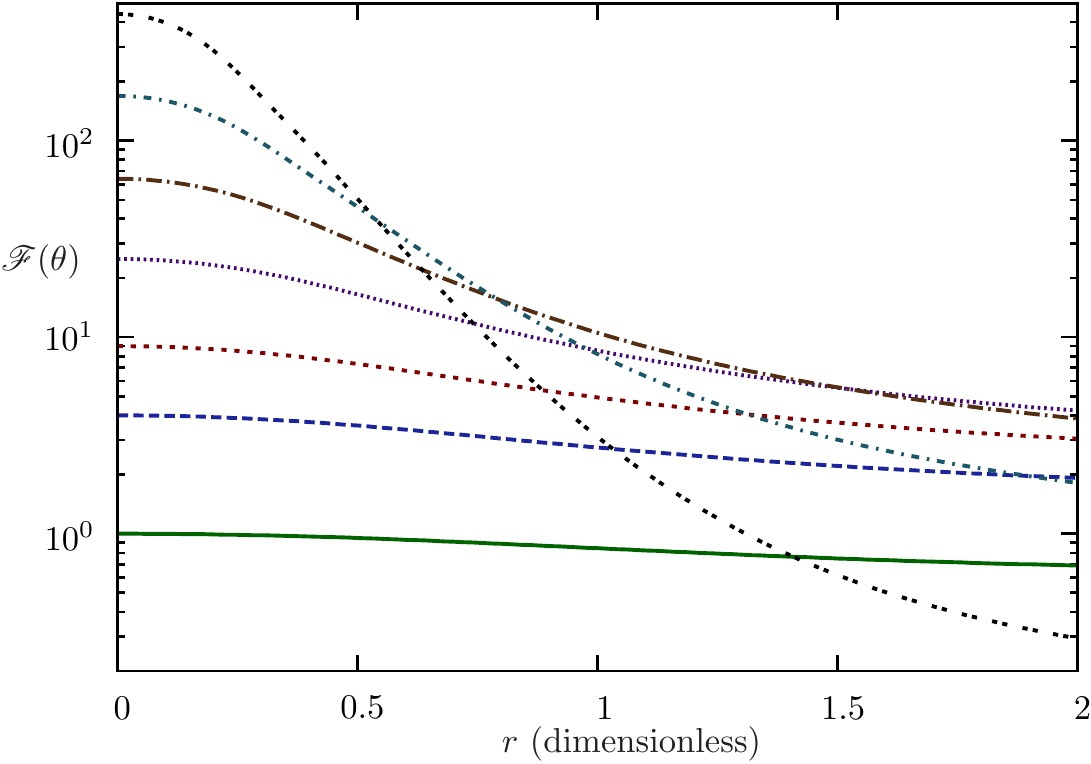}
  \caption[Single rail quantum Fisher information over $r$]{The quantum Fisher information $\mathscr{F}(\theta)$ over $r$ for Rob measuring the parameter $\theta$ when Alice has sent him the state using the single rail. The values of $N$ for each line are $1$, $2$, $3$, $5$, $8$, $13$ and $21$, corresponding to increasing Fisher information for $r=0$.}
  \label{fig:single-noon-r}
  \end{center}
\end{figure}

\begin{figure}[t]
  \begin{center}
      \includegraphics[width=\columnwidth]{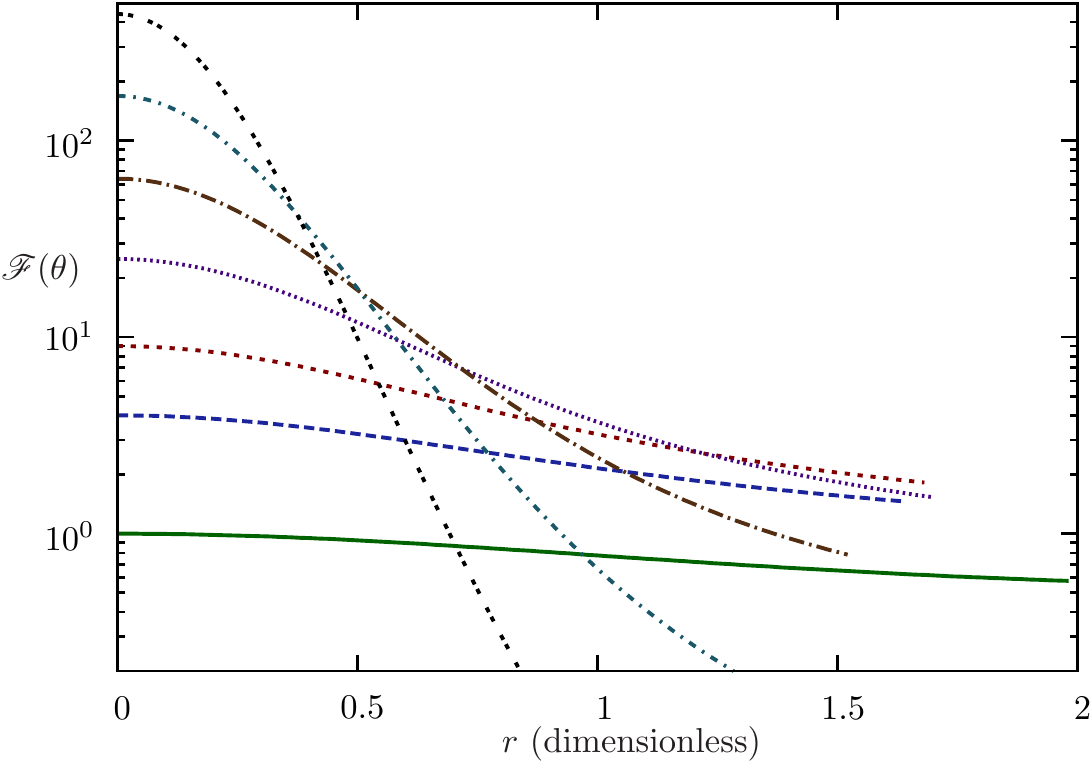}
  \caption[Dual rail quantum Fisher information over $N$]{The quantum Fisher information $\mathscr{F}(\theta)$ over $r$ for Rob measuring the parameter $\theta$ when Alice has sent him the state using the dual rail. The values of $N$ for each line are $1$, $2$, $3$, $5$, $8$, $13$ and $21$, corresponding to increasing Fisher information for $r=0$.}
  \label{fig:dual-noon-r}
  \end{center}
\end{figure}

The optimal value of $N$ is shown as a function of $r$ in Fig.\nobreakspace \ref {fig:single-noon-maxN} for the single- and dual-rail encodings. As $r\to 0$ the optimal $N$ diverges, since for the noiseless case ($r=0$) $\mathscr{F}(\theta)=N^2$, which grows without bound. For noiseless parameter estimation the NOON states are optimal \cite{Bollinger1996}, but in the presence of noise other states will give a higher Fisher information \cite{Maccone2008,Knysh2011}. We will explore these states in future work. For a nonzero squeezing parameter $r$, resulting in nonzero thermal noise, there is an exponential decay reducing the quantum Fisher information for larger $N$. To match this physical situation, we fitted this for each value of $r$ with a model of the form
\begin{equation}
  \label{eq:fitmodel}
  \mathscr{F}(\theta)= N^2 e^{- a(r) N + b(r)}.
\end{equation}
For large $r$, the coefficient $a(r)$ tended towards a linear function of $r$.
This function has a gradient of $(41.6 \pm 0.3) \times 10^{-3}$, when calculated using data for which $2.08\le r \le 3.10$.
It would be possible to use the linear dependence of the fitting parameter $a(r)$ to find the maximum value of $N$ for large $r$.
However, the interesting behaviour happens near $r=1$, because this is where the expected number of noise excitations is approximately $1$.

Finally, we show the quantum Fisher information is a function of increasing noise parameter $r$ for various values of $N$ in Figs.\nobreakspace \ref {fig:single-noon-r} and\nobreakspace  \ref {fig:dual-noon-r} for the single and dual rails, respectively.
Due to the large computational requirements of calculating these graphs, some of the lines stop before the end of the graph.
The single rail performs better than the dual rail for all values of $r$.

\section{Conclusions}
We studied the quantum Fisher information of single- and dual-rail NOON states sent from an inertial observer, Alice, to an accelerated observer, Rob.
The extractable information about the relative phase $\theta$ between the two terms in the NOON state is degraded due to the noise from the Unruh-Hawking effect.
The noise is an amplification channel, in that it depends on the number of excitations already present.
This $N$-dependent noise leads to an optimal $N$ for maximum information transfer, given a particular amount of noise, fixed by the squeezing parameter.
The quantum Fisher information degrades exponentially for increasing noise.
The single-rail encoding of the NOON states outperforms the dual-rail encoding, which is the opposite behaviour of earlier protocols that encoded the parameter in the amplitude, rather than the phase. 

\section{Acknowledgements}
DH would like to thank Michael Skotiniotis for useful discussions.

\bibliography{parameter-estimation.bib}

\end{document}